\documentclass[conference]{IEEEtran}
\IEEEoverridecommandlockouts
\usepackage{cite}
\usepackage{amsmath,amssymb,amsfonts}
\usepackage{algorithmic}
\usepackage{algorithm}
\usepackage{graphicx}
\usepackage{textcomp}
\usepackage{xcolor}
\usepackage{siunitx}
\usepackage{tikz}
\usepackage{pgfplots}
\usepgfplotslibrary{statistics}
\pgfplotsset{compat=1.8}
\usepackage{caption}
\newcommand{\smallplotheight}{0.75\columnwidth} % 0.75
\newcommand{\smallplotwidth}{0.55\columnwidth}
\newcommand{\normalplotheight}{0.7\columnwidth}
\newcommand{\normalplotwidth}{0.8\columnwidth}
\newcommand{\boxplotheight}{0.625\columnwidth} % 0.6
\newcommand{\boxplotwidth}{0.85\columnwidth}

\newcommand{\pltUniPowCov}{uni pow cov}
\newcommand{\pltHlsDLcb}{$\mbH_{\text{LS}}$, DL cb}
\newcommand{\pltHlsULcb}{$\mbH_{\text{LS}}$, UL cb}
\newcommand{\pltHompDLcb}{$\mbH_{\text{OMP}}$, DL cb}
\newcommand{\pltHompULcb}{$\mbH_{\text{OMP}}$, UL cb}
\newcommand{\pltHDLcb}{$\mbH$, DL cb}
\newcommand{\pltHULcb}{$\mbH$, UL cb}
\newcommand{\pltHlsWfCov}{$\mbH_{\text{LS}}$, wf cov}
\newcommand{\pltHompWfCov}{$\mbH_{\text{OMP}}$, wf cov}
\newcommand{\pltHeigsp}{$\mbH$, uni pow eigsp}
\newcommand{\pltWfCov}{$\mbH$, wf cov}

\newcommand{\lineWidth}{1.0pt}
\newcommand{\markSize}{2.0pt}
\definecolor{ourdarkblue}{RGB}{30, 100, 200}
\definecolor{ourdarkgreen}{RGB}{0, 100, 0}
\definecolor{ouryellow}{RGB}{220, 210, 50}
\tikzset{cbUL/.style={mark options={solid}, color=TUMBeamerOrange, line width=\lineWidth, mark=x, mark size=\markSize, solid}}
\tikzset{cbDL/.style={mark options={solid}, color=TUMBeamerOrange, line width=\lineWidth, mark=square, mark size=\markSize, dashed}}
\tikzset{cnnUL/.style={mark options={solid}, color=TUMBeamerRed, line width=\lineWidth, mark=o, mark size=\markSize, solid}}
\tikzset{cnnDL/.style={mark options={solid}, color=TUMBeamerRed, line width=\lineWidth, mark=triangle, mark size=\markSize, dashed}}
\tikzset{hlsUL/.style={mark options={solid}, color=TUMBeamerBlue, line width=\lineWidth, mark=star, mark size=\markSize, solid}}
\tikzset{hlsDL/.style={mark options={solid}, color=TUMBeamerBlue, line width=\lineWidth, mark=Mercedes, mark size=\markSize, dashed}}
\tikzset{hlsWF/.style={mark options={solid}, color=TUMBeamerBlue, line width=\lineWidth, mark=diamond, mark size=\markSize, dotted}}
\tikzset{hompUL/.style={mark options={solid}, color=black, line width=\lineWidth, mark=star, mark size=\markSize, solid}}
\tikzset{hompDL/.style={mark options={solid}, color=black, line width=\lineWidth, mark=Mercedes, mark size=\markSize, dashed}}
\tikzset{hompWF/.style={mark options={solid}, color=black, line width=\lineWidth, mark=diamond, mark size=\markSize, dotted}}
\tikzset{unipow/.style={mark options={solid}, color=ourdarkgreen, line width=\lineWidth, mark=diamond, mark size=\markSize, dotted}}

\newcommand{\whiskerstyle}{dashed}

\def\BibTeX{{\rm B\kern-.05em{\sc i\kern-.025em b}\kern-.08em
    T\kern-.1667em\lower.7ex\hbox{E}\kern-.125emX}}

\usepackage{mathtools}	% loads amsmath
\usepackage{amssymb}
\usepackage{amsthm}

\usepackage[utf8]{inputenc}

\usepackage{ifthen}

\usepackage{microtype}

\usepackage{caption}

\usepackage{bm}

\usepackage[draft,nomargin,marginclue,inline]{fixme}
\makeatletter
\renewcommand*\FXLayoutMarginClue[3]{%
  \marginpar[%
  \raggedleft\@fxuseface{margin}\textcolor{red}{\ignorespaces $ \Rightarrow $}]{%
    \raggedright\@fxuseface{margin}\textcolor{red}{\ignorespaces $ \Leftarrow $}}}
\makeatother	% change color and displayed text in margin
\usepackage{tumcolor}

\usepackage{pgfplotstable}	% loads pgfplots, tikz
\pgfplotsset{
	discard if/.style 2 args={
        x filter/.append code={
            \edef\tempa{\thisrow{#1}}
            \edef\tempb{#2}
            \ifx\tempa\tempb
                
            \fi
        }
    },
    discard if not/.style 2 args={
        x filter/.append code={
            \edef\tempa{\thisrow{#1}}
            \edef\tempb{#2}
            \ifx\tempa\tempb
            \else
                
            \fi
        }
    }
}

\usepackage{cite}

\usepackage[acronym,shortcuts]{glossaries}
\newacronym{cnn}{CNN}{convolutional neural network}
\newacronym{ula}{ULA}{uniform linear array}

\usepackage{cleveref}

\usepackage{enumitem}

\tikzset{algorithm1/.style={mark options={solid},color=TUMBeamerBlue, line width=\lineWidth, mark=square, dashed}}

\pgfplotscreateplotcyclelist{miko}{%
{color=TUMBeamerRed, dash pattern=on 5pt off 1pt on 1pt off 1pt, line width=1.5pt},%
{color=TUMBeamerOrange, dash pattern=on 5pt off 2pt, line width=1.5pt,
mark=o},%
{color=black, dash pattern=on 5pt off 2pt on 3pt off 2pt, line width=1.5pt},%
{color=TUMDarkerBlue, line width=1.5pt},%
{color=black, dotted, line width=1.5pt},%
{color=TUMBeamerGreen, dotted, line width=1.5pt},%
{color=TUMBeamerYellow, dotted, line width=1.5pt},%
{color=TUMBeamerDarkRed, dotted, line width=1.5pt},%
{color=TUMBeamerLightGreen, dotted, line width=1.5pt},%
{color=TUMIvory, dotted, line width=1.5pt},%
{color=TUMLightBlue, dotted, line width=1.5pt},%
}

\DeclareMathOperator*{\argmax}{arg\,max}

\DeclareMathOperator{\expec}{E}

\DeclareMathOperator{\tr}{tr}

\newcommand*{\C}{\mathbb{C}}

\newcommand{\herm}{{\operatorname{H}}}

\definecolor{myblue}{RGB}{30, 100, 200}

\newlength{\leftstackrelawd}
\newlength{\leftstackrelbwd}
\def\leftstackrel#1#2{\settowidth{\leftstackrelawd}%
	{${{}^{#1}}$}\settowidth{\leftstackrelbwd}{$#2$}%
	\addtolength{\leftstackrelawd}{-\leftstackrelbwd}%
	\leavevmode\ifthenelse{\lengthtest{\leftstackrelawd>0pt}}%
	{\kern-.5\leftstackrelawd}{}\mathrel{\mathop{#2}\limits^{#1}}}

\newcommand{\mbD}{\bm{D}}

\newcommand{\mbG}{\bm{G}}
\newcommand{\mbH}{\bm{H}}
\newcommand{\mbI}{\bm{I}}

\newcommand{\mbN}{\bm{N}}

\newcommand{\mbP}{\bm{P}}
\newcommand{\mbQ}{\bm{Q}}

\newcommand{\mbY}{\bm{Y}}

\newcommand{\mbg}{\bm{g}}
\newcommand{\mbh}{\bm{h}}

\newcommand{\mbn}{\bm{n}}

\newcommand{\mbs}{\bm{s}}
\newcommand{\mbt}{\bm{t}}

\newcommand{\mbx}{\bm{x}}
\newcommand{\mby}{\bm{y}}

\newcommand{\mbzero}{{\bm{0}}}

\Crefname{figure}{Fig.}{Figs.}

\newacronym{AWGN}{AWGN}{additive white Gaussian noise}
\newacronym{BLMMSE}{BLMMSE}{Bussgang LMMSE}
\newacronym{BS}{BS}{base station}
\newacronym{CNN}{CNN}{convolutional neural network}
\newacronym{CSI}{CSI}{channel state information}
\newacronym{CSIT}{CSIT}{channel state information at the transmitter}
\newacronym{DFT}{DFT}{discrete Fourier transform}
\newacronym{DL}{DL}{downlink}
\newacronym{DNN}{DNN}{deep neural network}
\newacronym{DoA}{DoA}{direction of arrival}
\newacronym{FDD}{FDD}{frequency division duplex}
\newacronym{LMMSE}{LMMSE}{linear minimum mean square error}
\newacronym{LOS}{LOS}{line of sight}
\newacronym{LS}{LS}{least squares}
\newacronym{MSE}{MSE}{mean squared error}
\newacronym{MIMO}{MIMO}{multiple-input multiple-output}
\newacronym{MPC}{MPC}{multi-path component}
\newacronym{MT}{MT}{mobile terminal}
\newacronym{NLOS}{NLOS}{non-line of sight}
\newacronym{NN}{NN}{neural network}
\newacronym{O2I}{O2I}{outdoor-to-indoor}
\newacronym{OMP}{OMP}{orthogonal matching pursuit}
\newacronym{PGD}{PGD}{projected gradient descent}
\newacronym{PSD}{PSD}{power spectral density}
\newacronym{SNR}{SNR}{signal-to-noise ratio}
\newacronym{TDD}{TDD}{time division duplex}
\newacronym{UL}{UL}{uplink}
\newacronym{ULA}{ULA}{uniform linear array}
\newacronym{UMa}{UMa}{urban macrocell}

\pgfplotsset{compat=1.15}

\newcommand{\Ncbentries}{K}
\newcommand{\Nrx}{N_{\mathrm{rx}}}
\newcommand{\Ntx}{N_{\mathrm{tx}}}
\newcommand{\Ttr}{T_{\mathrm{train}}}

\begin{document}

\title{Unsupervised Learning of Adaptive Codebooks for Deep Feedback Encoding in FDD Systems\thanks{\copyright This work has been submitted to the IEEE for possible publication. Copyright may be transferred without notice, after which this version may no longer be accessible.
}}

\author{\centerline{Nurettin Turan${^*}$, Michael Koller$^*$, Samer Bazzi$^\dagger$, Wen Xu$^\dagger$, and Wolfgang Utschick$^*$ }\\
\IEEEauthorblockA{$^*$Professur f\"ur Methoden der Signalverarbeitung, Technische Universit\"at M\"unchen, 80290 Munich, Germany\\ 
$^\dagger$Huawei Technologies Duesseldorf GmbH, 80992 Munich, Germany\\
Email: \small{\texttt{\{nurettin.turan,michael.koller,utschick\}@tum.de,\{samer.bazzi,wen.dr.xu\}@huawei.com}}}
}

\maketitle

\begin{abstract}
In this work, we propose a joint adaptive codebook construction and feedback generation scheme in \ac{FDD} systems.
Both unsupervised and supervised deep learning techniques are used for this purpose.
Based on a recently discovered equivalence of \ac{UL} and \ac{DL} \ac{CSI} in terms of neural network learning, the codebook and associated deep encoder for feedback signaling is based on \ac{UL} data only.
Subsequently, the feedback encoder can be offloaded to the \acp{MT} to generate channel feedback there as efficiently as possible, without any training effort at the terminals or corresponding transfer of training and codebook data.
Numerical simulations demonstrate the promising performance of the proposed method.
\end{abstract}


\section{Introduction}
Massive \ac{MIMO} technology is one of the most prominent directions to scale up capacity and throughput in modern communication systems \cite{marzetta}.
In particular, the multi-antennas support at the \ac{BS} makes simple techniques such as spatial multiplexing and beamforming very efficient regarding the spectrum or the bandwidth utilization.
In order to take full advantage of this technology, the \ac{BS} must have the best possible channel estimation.
However, considering the typically stringent delay requirements in wireless mobile communication systems, the \ac{CSI} has to be acquired in very short regular time intervals.
A variety of solution approaches developed for this purpose are based on \ac{TDD} mode.

On the other hand, in \ac{FDD} mode, the \ac{BS} and the \ac{MT} transmit in the same time slot but at different frequencies. This breaks the reciprocity between \ac{UL} \ac{CSI} and \ac{DL} \ac{CSI}, unlike in \ac{TDD} systems, and makes it difficult for the network operators with FDD licenses to obtain an accurate \ac{DL} \ac{CSI} estimate for transmit signal processing \cite{2019massive}. An obvious solution to the problem is to either extrapolate the \ac{DL} \ac{CSI} from the estimate of the \ac{UL} \ac{CSI} at the BS, or to transfer the \ac{DL} \ac{CSI} estimated at the \ac{MT} to the \ac{BS} directly or in a highly compressed version. However, the most common solution in practice is to avoid the direct feedback of the \ac{CSI} and use only selection indices that determine an element from a finite set of channel properties or from a finite set of predefined beamformer or precoder configurations. The latter method is also studied in this paper.

\begin{figure}[h]
\centering
\resizebox{0.475\textwidth}{!}{%

\tikzset{every picture/.style={line width=0.75pt}} %set default line width to 0.75pt        

\begin{tikzpicture}[x=0.75pt,y=0.75pt,yscale=-1,xscale=1]

%Curve Lines [id:da6463187973732616] 
\draw    (383,112) .. controls (348.35,132.79) and (333.3,87.91) .. (287.4,87.01) ;
\draw [shift={(286,87)}, rotate = 360] [color={rgb, 255:red, 0; green, 0; blue, 0 }  ][line width=0.75]    (10.93,-3.29) .. controls (6.95,-1.4) and (3.31,-0.3) .. (0,0) .. controls (3.31,0.3) and (6.95,1.4) .. (10.93,3.29)   ;
%Rounded Rect [id:dp24515175245695264] 
\draw  [fill={rgb, 255:red, 74; green, 144; blue, 226 }  ,fill opacity=0.3 ] (232,170) .. controls (232,165.58) and (235.58,162) .. (240,162) -- (270,162) .. controls (274.42,162) and (278,165.58) .. (278,170) -- (278,194) .. controls (278,198.42) and (274.42,202) .. (270,202) -- (240,202) .. controls (235.58,202) and (232,198.42) .. (232,194) -- cycle ;
%Straight Lines [id:da43670138328003816] 
\draw    (255,117) -- (255,148) ;
\draw [shift={(255,150)}, rotate = 270] [color={rgb, 255:red, 0; green, 0; blue, 0 }  ][line width=0.75]    (10.93,-3.29) .. controls (6.95,-1.4) and (3.31,-0.3) .. (0,0) .. controls (3.31,0.3) and (6.95,1.4) .. (10.93,3.29)   ;
%Rounded Rect [id:dp13195195497504164] 
\draw  [fill={rgb, 255:red, 245; green, 166; blue, 35 }  ,fill opacity=0.3 ] (79,73) .. controls (79,68.58) and (82.58,65) .. (87,65) -- (117,65) .. controls (121.42,65) and (125,68.58) .. (125,73) -- (125,97) .. controls (125,101.42) and (121.42,105) .. (117,105) -- (87,105) .. controls (82.58,105) and (79,101.42) .. (79,97) -- cycle ;
%Straight Lines [id:da03883995062390988] 
\draw    (222,87) -- (140,87) ;
\draw [shift={(138,87)}, rotate = 360] [color={rgb, 255:red, 0; green, 0; blue, 0 }  ][line width=0.75]    (10.93,-3.29) .. controls (6.95,-1.4) and (3.31,-0.3) .. (0,0) .. controls (3.31,0.3) and (6.95,1.4) .. (10.93,3.29)   ;
%Rounded Rect [id:dp33374729004828074] 
\draw  [color={rgb, 255:red, 128; green, 128; blue, 128 }  ,draw opacity=1 ] (354,221.4) .. controls (354,217.87) and (356.87,215) .. (360.4,215) -- (379.6,215) .. controls (383.13,215) and (386,217.87) .. (386,221.4) -- (386,259.6) .. controls (386,263.13) and (383.13,266) .. (379.6,266) -- (360.4,266) .. controls (356.87,266) and (354,263.13) .. (354,259.6) -- cycle ;
%Shape: Rectangle [id:dp2610722806330268] 
\draw  [color={rgb, 255:red, 128; green, 128; blue, 128 }  ,draw opacity=1 ] (357,221.4) -- (383,221.4) -- (383,259.6) -- (357,259.6) -- cycle ;
%Rounded Rect [id:dp018452198608421844] 
\draw  [color={rgb, 255:red, 128; green, 128; blue, 128 }  ,draw opacity=1 ] (405,78.4) .. controls (405,74.87) and (407.87,72) .. (411.4,72) -- (430.6,72) .. controls (434.13,72) and (437,74.87) .. (437,78.4) -- (437,116.6) .. controls (437,120.13) and (434.13,123) .. (430.6,123) -- (411.4,123) .. controls (407.87,123) and (405,120.13) .. (405,116.6) -- cycle ;
%Shape: Rectangle [id:dp3503267606845468] 
\draw  [color={rgb, 255:red, 128; green, 128; blue, 128 }  ,draw opacity=1 ] (408,78.4) -- (434,78.4) -- (434,116.6) -- (408,116.6) -- cycle ;
%Rounded Rect [id:dp22116465862733725] 
\draw  [color={rgb, 255:red, 128; green, 128; blue, 128 }  ,draw opacity=1 ] (454,91.4) .. controls (454,87.87) and (456.87,85) .. (460.4,85) -- (479.6,85) .. controls (483.13,85) and (486,87.87) .. (486,91.4) -- (486,129.6) .. controls (486,133.13) and (483.13,136) .. (479.6,136) -- (460.4,136) .. controls (456.87,136) and (454,133.13) .. (454,129.6) -- cycle ;
%Shape: Rectangle [id:dp9996015319018028] 
\draw  [color={rgb, 255:red, 128; green, 128; blue, 128 }  ,draw opacity=1 ] (457,91.4) -- (483,91.4) -- (483,129.6) -- (457,129.6) -- cycle ;
%Rounded Rect [id:dp10774378696214915] 
\draw  [color={rgb, 255:red, 128; green, 128; blue, 128 }  ,draw opacity=1 ] (419,139.4) .. controls (419,135.87) and (421.87,133) .. (425.4,133) -- (444.6,133) .. controls (448.13,133) and (451,135.87) .. (451,139.4) -- (451,177.6) .. controls (451,181.13) and (448.13,184) .. (444.6,184) -- (425.4,184) .. controls (421.87,184) and (419,181.13) .. (419,177.6) -- cycle ;
%Shape: Rectangle [id:dp3750016372905083] 
\draw  [color={rgb, 255:red, 128; green, 128; blue, 128 }  ,draw opacity=1 ] (422,139.4) -- (448,139.4) -- (448,177.6) -- (422,177.6) -- cycle ;
%Rounded Rect [id:dp6698569940641905] 
\draw  [fill={rgb, 255:red, 126; green, 211; blue, 33 }  ,fill opacity=0.3 ] (230,74) .. controls (230,69.58) and (233.58,66) .. (238,66) -- (268,66) .. controls (272.42,66) and (276,69.58) .. (276,74) -- (276,98) .. controls (276,102.42) and (272.42,106) .. (268,106) -- (238,106) .. controls (233.58,106) and (230,102.42) .. (230,98) -- cycle ;
%Curve Lines [id:da18168171902203967] 
\draw    (102,113) .. controls (102,129.66) and (213.41,141.52) .. (240.45,148.57) ;
\draw [shift={(242,149)}, rotate = 196.26] [color={rgb, 255:red, 0; green, 0; blue, 0 }  ][line width=0.75]    (10.93,-3.29) .. controls (6.95,-1.4) and (3.31,-0.3) .. (0,0) .. controls (3.31,0.3) and (6.95,1.4) .. (10.93,3.29)   ;
%Rounded Rect [id:dp6529714522995136] 
\draw  [fill={rgb, 255:red, 74; green, 144; blue, 226 }  ,fill opacity=0.3 ] (359,229) .. controls (359,227.34) and (360.34,226) .. (362,226) -- (378,226) .. controls (379.66,226) and (381,227.34) .. (381,229) -- (381,238) .. controls (381,239.66) and (379.66,241) .. (378,241) -- (362,241) .. controls (360.34,241) and (359,239.66) .. (359,238) -- cycle ;
%Straight Lines [id:da4170664338753678] 
\draw    (381,234) -- (406.33,234.31) ;
\draw [shift={(408.33,234.33)}, rotate = 180.7] [color={rgb, 255:red, 0; green, 0; blue, 0 }  ][line width=0.75]    (10.93,-3.29) .. controls (6.95,-1.4) and (3.31,-0.3) .. (0,0) .. controls (3.31,0.3) and (6.95,1.4) .. (10.93,3.29)   ;
%Rounded Rect [id:dp41670436611605366] 
\draw  [color={rgb, 255:red, 128; green, 128; blue, 128 }  ,draw opacity=1 ] (410.5,85.7) .. controls (410.5,83.93) and (411.93,82.5) .. (413.7,82.5) -- (428.3,82.5) .. controls (430.07,82.5) and (431.5,83.93) .. (431.5,85.7) -- (431.5,95.3) .. controls (431.5,97.07) and (430.07,98.5) .. (428.3,98.5) -- (413.7,98.5) .. controls (411.93,98.5) and (410.5,97.07) .. (410.5,95.3) -- cycle ;
%Rounded Rect [id:dp4161122351822395] 
\draw  [color={rgb, 255:red, 128; green, 128; blue, 128 }  ,draw opacity=1 ] (459.5,98.7) .. controls (459.5,96.93) and (460.93,95.5) .. (462.7,95.5) -- (477.3,95.5) .. controls (479.07,95.5) and (480.5,96.93) .. (480.5,98.7) -- (480.5,108.3) .. controls (480.5,110.07) and (479.07,111.5) .. (477.3,111.5) -- (462.7,111.5) .. controls (460.93,111.5) and (459.5,110.07) .. (459.5,108.3) -- cycle ;
%Rounded Rect [id:dp6150812915155415] 
\draw  [color={rgb, 255:red, 128; green, 128; blue, 128 }  ,draw opacity=1 ] (424.5,147.7) .. controls (424.5,145.93) and (425.93,144.5) .. (427.7,144.5) -- (442.3,144.5) .. controls (444.07,144.5) and (445.5,145.93) .. (445.5,147.7) -- (445.5,157.3) .. controls (445.5,159.07) and (444.07,160.5) .. (442.3,160.5) -- (427.7,160.5) .. controls (425.93,160.5) and (424.5,159.07) .. (424.5,157.3) -- cycle ;
%Curve Lines [id:da27281559801222] 
\draw    (283,181) .. controls (326.5,182.5) and (346,208) .. (362,226) ;
%Straight Lines [id:da8983713855112414] 
\draw    (378,241) -- (386.14,251.05) ;
\draw [shift={(387.4,252.6)}, rotate = 230.98] [color={rgb, 255:red, 0; green, 0; blue, 0 }  ][line width=0.75]    (10.93,-3.29) .. controls (6.95,-1.4) and (3.31,-0.3) .. (0,0) .. controls (3.31,0.3) and (6.95,1.4) .. (10.93,3.29)   ;
%Straight Lines [id:da2626912041267011] 
\draw    (326.33,233.33) -- (356.67,233.65) ;
\draw [shift={(358.67,233.67)}, rotate = 180.59] [color={rgb, 255:red, 0; green, 0; blue, 0 }  ][line width=0.75]    (10.93,-3.29) .. controls (6.95,-1.4) and (3.31,-0.3) .. (0,0) .. controls (3.31,0.3) and (6.95,1.4) .. (10.93,3.29)   ;

% Text Node
\draw (426,49) node [anchor=north west][inner sep=0.75pt]   [align=left] {MT};
% Text Node
\draw (325,123) node [anchor=north west][inner sep=0.75pt]   [align=left] %{$\mbH_{n,\text{UL}}$};
{$\mathcal{H}^{\text{UL}}$};
% Text Node
\draw (168,116) node [anchor=north west][inner sep=0.75pt]   [align=left] {learn $f_\text{DNN}(\cdot)$};
% Text Node
\draw (238,174) node [anchor=north west][inner sep=0.75pt]   [align=left] {DNN};
% Text Node
\draw (296,63) node [anchor=north west][inner sep=0.75pt]   [align=left] {collect UL data};
% Text Node
\draw (95,77) node [anchor=north west][inner sep=0.75pt]   [align=left] {$\mathcal{Q}$};
% Text Node
\draw (150,61.5) node [anchor=north west][inner sep=0.75pt]   [align=left] {construct $\mathcal{Q}$};
% Text Node
\draw (242,78) node [anchor=north west][inner sep=0.75pt]   [align=left] {BS};
% Text Node
\draw (361,231.6) node [anchor=north west][inner sep=0.75pt]  [font=\tiny] [align=left] {DNN};
% Text Node
\draw (412.67,228.67) node [anchor=north west][inner sep=0.75pt]   [align=left] {feedback index $k^\star$};
% Text Node
\draw (292.79,155.97) node [anchor=north west][inner sep=0.75pt]  [rotate=-27.38] [align=left] {share parameters};
% Text Node
\draw (240.17,228.67) node [anchor=north west][inner sep=0.75pt]   [align=left] {observation $\mbY$};

\end{tikzpicture}

}
    \caption{Structure of the proposed approach with codebook construction and learning of the DNN classifier at the BS, which is then subsequently offloaded to the MT.}
    \label{fig:my_label}
\end{figure}
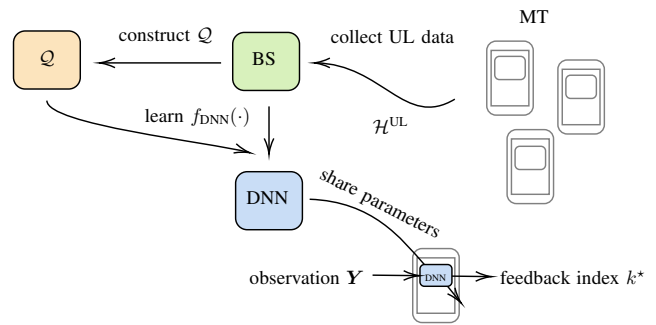

In particular, we propose to construct a finite codebook $ \mathcal{Q} $ of different precoder configurations at the \ac{BS} using an unsupervised learning method.
Although the codebook covers precoding matrices for the \ac{DL} operation, the construction of the codebook is solely based on a set of \ac{UL} channels $\mathcal{H}^{\text{UL}}$ which have been collected at the BS.
Following the codebook construction, a \ac{DNN} classifier $ f_\text{DNN}(\cdot) $ is trained at the \ac{BS} again solely based on the \ac{UL} samples, which assigns the index $ k^\star $ of the most appropriate precoding matrix to a given \ac{DL} observation.
The exclusive use of \ac{UL} data for the training of function blocks, which are exclusively intended for the \ac{DL} channel, follows the recent results in \cite{utschick2021,rizzello2021}. The proposed approach is then to subsequently offload this deep feedback encoder to each \ac{MT} in the cell. See \Cref{fig:my_label} for a sketch of the proposed overall concept.

Thus, the core idea of our scheme is that the neural network encoder trained on \ac{UL} data at the \ac{BS} can be applied to \ac{DL} data without any further adaptation from any mobile device to which the encoder is offloaded. Training on the \ac{MT} is no longer necessary at all, making it possible to quickly update the encoder on the \ac{MT} at any time and place with an updated version of the codebook, e.g., when moving from one cell to another or for different locations of the \ac{MT} in the cell.
Based on the presented simulation results, we are able to demonstrate the promising performance of the proposed technique.

\section{System Model}
The \ac{DL} received signal of a single user \ac{MIMO} system can be expressed as
$\mby = \mbH \mbx + \mbn$, where $\mby \in \C^{\Nrx}$ is the receive vector, $\mbx \in \C^{\Ntx}$ is the transmit vector sent over the \ac{MIMO} channel $\mbH \in \C^{\Nrx \times \Ntx}$, and $\mbn \sim \mathcal{N}_\C(\mathbf{0},% \mbC_{\mbn} =
\sigma_n^2 \mbI_{\Nrx})$ denotes the \ac{AWGN}.
In this paper, we consider system configurations with $\Nrx < \Ntx$. If both the transmitter and receiver know the channel perfectly, and assuming input data with Gaussian distribution, %Gaussian codebooks,
the capacity of the \ac{MIMO} channel is \cite{Goldsmith, Goldsmith2}:
\begin{equation}
    C = \max_{\mbQ \succeq \mbzero, \tr\mbQ\leq\rho} \log_2 \det\left( \mbI + \frac{1}{\sigma_n^2} \mbH \mbQ \mbH^\herm\right),
\end{equation}
where $\mbQ \in \C^{\Ntx \times \Ntx}$ is the transmit covariance matrix.
This assumes a transmit vector given by \( \mbx = \mbQ^{1/2} \mbs \) with \( \expec[\mbs \mbs^\herm] = \mbI_{\Ntx} \)~\cite{Love}. % "codeword" entfernt
%From an encoding perspective, the transmit vector $\mbx = \mbQ^{1/2} \mbs$, where the codeword $\mbs$ has the property \cite{Love}: $E[\mbs\mbs^\herm] = \mbI_{\Ntx}$.
The capacity-achieving transmit covariance matrix $\mbQ^\star$ of the link between the \ac{BS} and a \ac{MT} can be found by decomposing the channel into $\Nrx$ parallel streams and employing water-filling \cite{Telatar99capacityof}. 

%In \ac{FDD} systems, the downlink and the uplink channels are highly uncorrelated, because they are separated in frequency \cite{Love}. 
%Thus, no channel reciprocity can be assumed. 
In \ac{FDD} systems, channel reciprocity can generally not be assumed, e.g., \cite{Love}.
For this reason, only the \ac{MT} could compute the optimal transmit covariance matrix \( \mbQ^\star \) if it estimated the \ac{DL} \ac{CSI}. 
This makes some form of feedback from the \ac{MT} to the \ac{BS} necessary.
Ideally, the user would feed the complete \ac{DL} \ac{CSI} back to the \ac{BS}, which in general is considered to be infeasible.
Instead, a low-rate feedback link is used to transmit a small number of \( M \) bits back to the \ac{BS}.

Typically, the \( M \) feedback bits are used for encoding an index into a set of covariance matrices.
That is, the \ac{MT} and \ac{BS} share a {codebook} $ \mathcal{Q} = \{\mbQ_1, \mbQ_2, \dots, \mbQ_{K} \} $ of $ K = 2^M $ pre-computed transmit covariance matrices,
the \ac{MT} is assumed to estimate the \ac{DL} channel \( \mbH \) and then uses it to determine the best codebook entry \( \mbQ_{k^\star} \) in terms of maximum achievable rate, i.e.,
\begin{equation}\label{eq:codebook_index_selection}
    k^\star = \argmax_{k \in \{1, \dots, \Ncbentries\}} \log_2 \det\left( \mbI + \frac{1}{\sigma_n^2} \mbH \mbQ_k \mbH^\herm\right),
\end{equation}
or other practically useful selection criteria.
Finally, the feedback consists of the index \( k^\star \) encoded by $ M $ bits, and the \ac{BS} employs the transmit covariance matrix \( \mbQ_{k^\star} \) for data transmission.
\Cref{sec:codebook_design} describes an algorithm to obtain a codebook \( \mathcal{Q} \).

%\textcolor{red}{Weglassen? The \ac{MS} would have to send $\mbQ^*$ (or the singular vectors and the corresponding singular values) to the \ac{BS}. This would result in a huge feedback overhead.}
%Nevertheless, the \ac{BS} needs some sort of \ac{CSI}, referred to as \ac{CSIT}.
%The \ac{CSIT} can be acquired by using feedback, where a low-rate reverse link is used to provide information of the instantaneous downlink channel to the \ac{BS}. 

%We consider covariance quantization, where a codebook $\mathcal{Q} = \{\mbQ_1, \mbQ_2, \dots, \mbQ_{\Ncbentries=2^M}  \}$ is shared between the \ac{BS} and the \ac{MS}, where we have in total $N$ codebook entries and the amount of feedback bits is $M$.
%The \ac{MS} computes
%\begin{align*}
%    r_k(\mbH, \mbQ_k) &= \log_2 \det\left( \mbI + \frac{1}{\sigma_n^2} \mbH \mbQ_k %\mbH^\herm\right),\\
%    k^* &= \argmax_{k \in \{1, \cdots, N\}} r_k(\mbH, \mbQ_k),
%\end{align*}
%$\forall k \in \{1, \cdots, N\}$, and reports the index $k^*$ of the covariance matrix, which leads to the maximum rate $r^*_k$, to the BS.\newline

\section{Channel Model  and Data Generation}

We consider an \ac{FDD} system below $\SI{6}{\giga\hertz}$ and assume a frequency gap of $\SI{200}{\mega\hertz}$ between \ac{UL} and \ac{DL}.
Version $2.2$ of the QuaDRiGa channel simulator \cite{QuaDRiGa1, QuaDRiGa2} is used to generate \ac{CSI} for the \ac{UL} and \ac{DL} scenarios.

We simulate two \ac{UMa} single carrier scenarios: one with $(\Ntx, \Nrx) = (16, 4)$ and one with $ (\Ntx, \Nrx) = (32, 16)$.
In both cases, the \ac{UL} carrier frequency is $\SI{2.53}{\giga\hertz}$ and the \ac{DL} carrier frequency is $\SI{2.73}{\giga\hertz}$. 
The \ac{BS} is equipped with a \ac{ULA} with ``3GPP-3D'' antennas, and the \ac{MT} consists of a \ac{ULA} assuming ``omni-directional'' antennas. 
The \ac{BS} is placed at a height of $\SI{25}{\meter}$ and covers a sector of $\SI{120}{\degree}$, where the minimum distance of the \ac{MT} location to the \ac{BS} is $\SI{35}{\meter}$ and the maximum distance to the \ac{BS} is $\SI{500}{\meter}$. 
In $80\%$ of the cases, the \ac{MT} is located indoors at different floor levels, and in the case of outdoor locations the \ac{MT}'s height is $\SI{1.5}{\meter}$ in accordance with \cite{3gpp}.

As outlined in \cite{QuaDRiGa1}, many parameters such as path-loss, delay, and angular spreads, path-powers for each subpath, and antenna patterns are different in the \ac{UL} and \ac{DL} domain. 
However, the following parameters are identical in the \ac{UL} and \ac{DL} domain: \ac{BS} location and the \ac{MT} locations, propagation cluster delays and angles for each \ac{MPC}, and the spatial consistency of the large scale fading parameters. 
QuaDRiGa models \ac{MIMO} channels as
\begin{equation} 
\mbH = \sum_{\ell=1}^{L} \mbG_{\ell} e^{-2\pi j f_c \tau_{\ell}}, 
\end{equation}
where $\ell$ is the path number, and the number of \acp{MPC} $L$ depends on whether there is \ac{LOS}, \ac{NLOS}, or \ac{O2I} propagation: $L_\text{LOS} = 37$, $L_\text{NLOS} = 61$ or $L_\text{O2I} = 37$.
The carrier frequency is denoted by $f_c$ and the $\ell$-th path delay by $\tau_{\ell}$. 
The MIMO coefficient matrix $\mbG_{\ell}$ consists of one complex entry for each antenna pair, which comprises the attenuation of a path, the antenna radiation pattern weighting, and the polarization \cite{Kurras}.
The data is not post-processed: we work with channels including the path gain.

We generate datasets with $30\times 10^3$ channels for both the \ac{UL} and \ac{DL} of each scenario.
Defining a \ac{SNR} as $\text{SNR} = {E[\|\mbH \mbx \|^2_2]}/{E[\|\mbn \|^2_2]}$ and assuming  \( \mbQ = \frac{\rho}{\Ntx} \mbI \) (i.e., uniform power allocation), it holds: $\text{SNR} = \frac{\rho /\Ntx ||\mbH ||_F^2}{\sigma_n^2 \Nrx}$. 
In our simulations, we have a noise variance of $\sigma_n^2 = \SI{-114}{dBm}$ and a transmit power of $\rho = \SI{15}{dBm}$. 
We then select all \ac{DL} channels together with the corresponding \ac{UL} channels (with the same \ac{MT} locations) which fall within the \ac{SNR} range of $[\SI{-10}{\dB},\SI{20}{dB}]$ to constitute our datasets.
This leads to the following pairs of \ac{UL} and \ac{DL} datasets for the two considered scenarios:
\begin{align}
\mathcal{H}_{4\times16}^{\text{UL}}, \mathcal{H}_{4\times16}^{\text{DL}}, \mathcal{H}_{16\times32}^{\text{UL}}, \ \text{and} \ \mathcal{H}_{16\times32}^{\text{DL}},
\end{align}
with cardinalities $|\mathcal{H}_{4\times16}^{\text{UL}}|$ = $|\mathcal{H}_{4\times16}^{\text{DL}}| = 18773$ and $|\mathcal{H}_{16\times32}^{\text{UL}}| = |\mathcal{H}_{16\times32}^{\text{DL}}| = 16148$.

The \ac{UL} channels have a dimension of $16\times4$ or $32\times16$, respectively, depending on the scenario. The sets $\mathcal{H}_{4\times16}^{\text{UL}}$ and $\mathcal{H}_{16\times32}^{\text{UL}}$ contain transposed versions of the respective channels, i.e, with dimensions $4\times16$ or $16\times32$.

%\red{Overall, we generate the following pairs of \ac{UL} and \ac{DL} datasets for the two antenna configurations: $\mathcal{H}_{4\times16}^{\text{UL}}$, $\mathcal{H}_{4\times16}^{\text{DL}}$, $\mathcal{H}_{16\times32}^{\text{UL}}$, and $\mathcal{H}_{16\times32}^{\text{DL}}$, where $|\mathcal{H}_{4\times16}^{\text{UL}}|$ = $|\mathcal{H}_{4\times16}^{\text{DL}}| = 18773$ and $|\mathcal{H}_{16\times32}^{\text{UL}}|$, and $|\mathcal{H}_{16\times32}^{\text{DL}}| = 16148$.
%In our simulations the noise variance $\sigma_n^2 = \SI{-114}{dBm}$ and the transmit power $\rho = \SI{15}{dBm}$.}

\section{Unsupervised Codebook Design} % by using \ac{UL} Data}}
\label{sec:codebook_design}

In this section, we first briefly describe how we can construct a codebook given a set of training channels.
Then, we show that we can learn a codebook for the \ac{DL} domain by using a training set solely consisting of \ac{UL} channel data.

\subsection{Iterative Llyod Clustering Algorithm}
\label{sec:cb_design}
Let $ \mathcal{H} = \{ \mbH_n \}_{n=1}^{\Ttr} $ be a set of collected training channel matrices.
The goal is to obtain a corresponding codebook
$ \mathcal{Q} = \{ \mbQ_k \}_{k=1}^{\Ncbentries} $ of $ \Ncbentries = 2^M $ transmit covariance matrices.
To this end, we employ an iterative Lloyd clustering algorithm which alternates between two stages until a convergence criterion is met \cite{LiBuGr80}.
We write $ \{ \mbQ_k^{(i)} \}_{k=1}^{\Ncbentries} $ for the codebook in iteration $ i $.
Further, we define the spectral efficiency
\begin{equation}
    r(\mbH, \mbQ) = \log_2 \det\left( \mbI + \frac{1}{\sigma_n^2} \mbH \mbQ \mbH^\herm\right)
    \label{speceff}
\end{equation}
to be the cost criterion of interest.
Then, the two stages of iteration $ i $ are expressed as follows:
\begin{enumerate}
    \item Divide the training set $ \mathcal{H} $ into $ \Ncbentries $ clusters $ \mathcal{V}_k^{(i)} $:
    \begin{equation}\label{eq:lloyd_stage_1}
        \hspace*{-4mm}\mathcal{V}_k^{(i)} = \{ \mbH \in \mathcal{H} \mid r(\mbH, \mbQ_k^{(i)}) \geq r(\mbH, \mbQ_j^{(i)}), k\neq j \}.
    \end{equation}
    \item Find new covariance matrices or update the so called ``cluster centers'':
    \begin{align}\label{eq:lloyd_stage_2}
        &\mbQ_k^{(i+1)} = \argmax_{\mbQ \succeq \mbzero} \frac{1}{|\mathcal{V}_k^{(i)}|} \sum_{\mbH\in\mathcal{V}_k^{(i)}} r(\mbH,\mbQ) \\
        & \text{subject to} \quad \operatorname{trace}(\mbQ) \leq \rho \quad \text{and} \quad
        \operatorname{rank}\mbQ \leq \Nrx. \nonumber
    \end{align}
\end{enumerate}
%\wout{where}$ \mathcal{C} $ contains further covariance matrix constraints:
% \begin{equation}
%     \mathcal{C} & =
%     \left\{
%         \mbQ \in \C^{\Ntx \times \Ntx} \mid
%         \operatorname{tr}(\mbQ) \leq \rho,\,
%         \operatorname{rk}(\mbQ) \leq \Nrx)
%     \right\}.
% \end{equation}
\noindent Whereas~\eqref{eq:lloyd_stage_1} is easily obtained by computing the spectral efficiency \eqref{speceff} of each training channel matrix with every codebook element,
we solve the optimization problem~\eqref{eq:lloyd_stage_2} with \ac{PGD} as it can be found in~\cite{HuScJoUt08}.
\ac{PGD} starts from an initialization---in our case it is the scaled identity matrix---and updates  $ \mbQ $ using an unconstrained gradient step:
\begin{align}
    \mbQ \leftarrow \mbQ + \alpha \mbg_{\mbQ}.
\end{align}
The gradient $ \mbg_{\mbQ} $ of the objective function in~\eqref{eq:lloyd_stage_2} with respect to $ \mbQ $, namely
\begin{equation}
    \mbg_{\mbQ} = \frac{1}{\sigma_n^2 \ln(2)} \sum_{\mbH \in \mathcal{V}_k^{(i)}} \mbH^\herm \left( \mbI + \frac{1}{\sigma_n^2} \mbH \mbQ \mbH^\herm \right)^{-1} \mbH,
\end{equation}
leaves the updated $\mbQ$ still positive-semidefinite for arbitrary step sizes $ \alpha \geq 0 $.
The other constraints in \eqref{eq:lloyd_stage_2} are thereafter enforced via a projection step of $\mbQ$ onto the constraint set.
The projection is done via water-filling.
The two stages consisting of unconstrained gradient update followed by projection are iterated until convergence.
In our simulations, an Armijo rule controls the step size.
More details about this version of \ac{PGD} can be found in~\cite{HuScJoUt08}.

\subsection{Codebook Construction---UL versus DL Data}

We split the two sets $\mathcal{H}_{4\times16}^{\text{UL}}$, $\mathcal{H}_{4\times16}^{\text{DL}}$ into a training set with $\Ttr = 10^4$ samples, and the remaining samples constitute an evaluation set:
\begin{align}
\mathcal{H}_{4\times16}^{\text{UL,train}}, \mathcal{H}_{4\times16}^{\text{UL,eval}}, \mathcal{H}_{4\times16}^{\text{DL,train}}, \ \text{and} \ \mathcal{H}_{4\times16}^{\text{DL,eval}}.
\end{align}
However, the \ac{UL} evaluation set $\mathcal{H}_{4\times16}^{\text{UL,eval}}$ is not relevant for our considerations and the following transmit strategies are always evaluated on \( \mathcal{H}_{4\times 16}^{\text{DL,eval}} \).
%we do never use the set $\mathcal{H}_{4\times16}^{\text{UL,eval}}$.

\begin{figure}[tb]
    \centering
\begin{tikzpicture}
\begin{axis}[
  legend style={at={(0,1)},anchor=north west},
  boxplot/draw direction=y,
  ylabel={\scriptsize{spectral efficiency [bits/c.u.]}},
  height=\boxplotheight,
  width=\boxplotwidth,
  % ... and it means that we should describe intervals:
  %extra y ticks={5,15,25},
  xtick={1,2,...,50},
  %x tick label as interval,
  xticklabels={%
      \scriptsize{\pltUniPowCov},%
      \scriptsize{$M=3$, DL},%
      \scriptsize{$M=3$, UL},%
      \scriptsize{$M=6$, DL},%
      \scriptsize{$M=6$, UL},%
      \scriptsize{$M=12$, DL},%
      \scriptsize{$M=12$, UL},%
      \scriptsize{all train data DL},%
      \scriptsize{all train data UL},%
      \scriptsize{\pltHeigsp},%
      \scriptsize{\pltWfCov},%
  },
  x tick label style={
      text width=1.9cm,
      font=\footnotesize,
      align=right,
      rotate=80
  },
  xtick style={draw=none},
  y tick label style={
      font=\footnotesize,
      align=center,
  },
  y label style={at={(-0.08,0.5)}},
  cycle list={},
]
%\addplot [domain=1.5:19.5, thick, cyan] {11.128912417815629}; % median of M=3

% uniform power allocation
\addplot+[
    TUMBeamerRed,
    boxplot prepared={
      median=8.349689860148676,
      upper quartile=14.612202291480003,
      lower quartile=3.2180369215243636,
      upper whisker=26.584417675500962,
      lower whisker=0.3489717488209078,
      every whisker/.style={\whiskerstyle},
      every median/.style={solid},
      every box/.style={fill=TUMBeamerRed,opacity = 0.25}
    },
    ]
    coordinates {};

% cb DL 3
\addplot+[
    TUMBlue,
    boxplot prepared={
      lower whisker=0.2173913025507126,
      lower quartile=5.399506789641115,
      median=11.128912417815629,
      upper quartile=17.787247778551805,
      upper whisker=32.683300160249864,
      every whisker/.style={\whiskerstyle},
      every median/.style={solid},
      every box/.style={fill=TUMBlue,opacity = 0.2}
    },
    ]
    coordinates {};
    
% cb UL 3
\addplot+[
    TUMBeamerGreen,
    %dashed,
    boxplot prepared={
      lower whisker=0.33035014669437845,
      lower quartile=5.382910723941591,
      median=11.152375644005158,
      upper quartile=17.77067425414934,
      upper whisker=32.07315141852086,
      every whisker/.style={\whiskerstyle},
      every median/.style={solid},
      every box/.style={fill=TUMBeamerGreen,opacity = 0.2}
    },
    ]
    coordinates {};

% DL 6
\addplot+[
    TUMBlue,
    boxplot prepared={
      lower whisker=0.681222890684834,
      lower quartile=5.754131031791986,
      median=11.777991017625055,
      upper quartile=18.641529315773727,
      upper whisker=32.66206098120042,
      every whisker/.style={\whiskerstyle},
      every median/.style={solid},
      every box/.style={fill=TUMBlue,opacity = 0.4}
    },
    ]
    coordinates {};
 
% UL 6  
\addplot+[
    TUMBeamerGreen,
    boxplot prepared={
      lower whisker=0.6617899174777325,
      lower quartile=5.703155972937565,
      median=11.779116671236295,
      upper quartile=18.60336720374872,
      upper whisker=32.64132252121264,
      every whisker/.style={\whiskerstyle},
      every median/.style={solid},
      every box/.style={fill=TUMBeamerGreen,opacity = 0.4}
    },
    ]
    coordinates {};

% DL 12
\addplot+[
    TUMBlue,
    boxplot prepared={
      lower whisker=0.9095790248346872,
      lower quartile=6.097795122759394,
      median=12.231885417543952,
      upper quartile=19.096235132449966,
      upper whisker=32.59094323788072,
      every whisker/.style={\whiskerstyle},
      every median/.style={solid},
      every box/.style={fill=TUMBlue,opacity = 0.7}
    },
    ]
    coordinates {};
 
% UL 12  
\addplot+[
    TUMBeamerGreen,
    boxplot prepared={
      lower whisker=0.9011135659665744,
      lower quartile=6.1132124122741205,
      median=12.230519535214302,
      upper quartile=19.09314449283985,
      upper whisker=32.45025549738799,
      every whisker/.style={\whiskerstyle},
      every median/.style={solid},
      every box/.style={fill=TUMBeamerGreen,opacity = 0.7}
    },
    ]
    coordinates {};

% wf cb DL
\addplot+[
    TUMBeamerOrange,
    boxplot prepared={
      lower whisker=1.0815172993305,
      lower quartile=6.215036748704905,
      median=12.307413552972724,
      upper quartile=19.218420906964713,
      upper whisker=32.34534020224435,
      every whisker/.style={\whiskerstyle},
      every median/.style={solid},
      every box/.style={fill=TUMBeamerOrange,opacity = 0.2}
    },
    ]
    coordinates {};

% wf cb UL
\addplot+[
    TUMBeamerOrange,
    boxplot prepared={
      lower whisker=0.9491484496560828,
      lower quartile=6.223272718606784,
      median=12.33864117921653,
      upper quartile=19.22687260315766,
      upper whisker=32.25715555790234,
      every whisker/.style={\whiskerstyle},
      every median/.style={solid},
      every box/.style={fill=TUMBeamerOrange,opacity = 0.7}
    },
    ]
    coordinates {};

% uni pow H
\addplot+[
    TUMDarkGray,
    boxplot prepared={
      lower whisker=1.14042986285615,
      lower quartile=7.147132574728896,
      median=14.523214477953129,
      upper quartile=21.892799160837,
      upper whisker=34.5327375451216,
      every whisker/.style={\whiskerstyle},
      every median/.style={solid},
      every box/.style={fill=TUMDarkGray,opacity = 0.25}
    },
    ]
    coordinates {};

% H wf
\addplot+[
    TUMDarkGray,
    boxplot prepared={
      lower whisker=1.758560081435993,
      lower quartile=7.746208293061338,
      median=14.662980867253147,
      upper quartile=21.90918050179577,
      upper whisker=34.532747040316906,
      every whisker/.style={\whiskerstyle},
      every median/.style={solid},
      every box/.style={fill=TUMDarkGray,opacity = 0.7}
    },
    ]
    coordinates {};

% means
\addplot[line width=1,solid, mark=*, opacity=0.5, mark size=0.5pt]%,mark=square]
        table[x=x,y=mean] {16x4_3_6_12bit_se_means.txt};
%        \addlegendentry{\footnotesize{mean}};

\legend{,,,,,,,,,,,\scriptsize{mean}};

\end{axis}
\end{tikzpicture}
\vspace{-0.2cm}
    \caption{Spectral efficiencies corresponding to codebooks of different sizes \( M \), constructed using either UL or DL training data.}
    \label{fig:16x4_Mbit_se}
\end{figure}
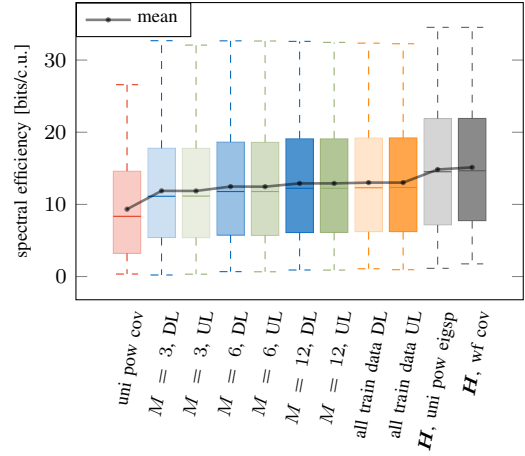

Since the distances of the \ac{MT} locations to the \ac{BS} range from $\SI{35}{m}$ to $\SI{500}{m}$ and since we cover an \ac{SNR} range of $[\SI{-10}{\dB},\SI{20}{dB}]$, the quality of the channels varies greatly.
For this reason, we display the distribution of the spectral efficiencies of the channels in the set $\mathcal{H}_{4\times16}^{\text{DL,eval}}$ for various transmit strategies via box plots. Every boxplot in \Cref{fig:16x4_Mbit_se} highlights the median, the first, and the third quartile, and the whiskers represent an interquartile range of 1.5.\\
\indent\emph{i}) The boxplot labled ``{\tt uni pow cov}'' represents uniform power allocation where the transmit covariance matrix is simply given by \( \mbQ = \frac{\rho}{\Ntx} \mbI \). 
In this case, no channel knowledge or codebook is used. \\
%We depict the ``uni pow'' transmit strategy, which does not require any knowledge on the channel or a codebook and  simply uses the weighted identity matrix as transmit covariance matrix, i.e., $\mbQ = \frac{\rho}{\Ntx}\mbI$.
%Note, that in this case $\Ntx$ data streams are transmitted, but ideally the \ac{MS} should receive up to $\Nrx$ streams.
%Nevertheless, the ``uni pow'' approach provides a lower bound on the spectral efficiencies.
\indent\emph{ii}) As a performance upper bound, \Cref{fig:16x4_Mbit_se} shows the spectral efficiency distribution obtained using the optimal transmit strategy, where the capacity-achieving transmit covariance matrix is computed for every channel via water-filling (``$\mbH$, {\tt wf cov}'').\\
%where in this case, for each channel the optimal (capacity-achieving) transmit covariance matrix is calculated.
\indent \emph{iii}) Moreover, \Cref{fig:16x4_Mbit_se} depicts the ``$\mbH$, {\tt uni pow eigsp}'' transmit strategy, where a transmit covariance matrix is calculated by allocating equal power on the eigenvectors of the channel.
That is, the channel is decomposed into $\Nrx$ parallel streams and $\frac{\rho}{\Nrx}$ power is allocated on each stream.
Note that this and the optimal approach are infeasible because the \ac{BS} would require full knowledge of the \ac{DL} channel. \\
\indent\emph{iv}) Furthermore, \Cref{fig:16x4_Mbit_se} contains codebook based transmit strategies with $M \in \{3,6,12\}$ bit. Recall, that a feedback index is selected via~\eqref{eq:codebook_index_selection} using the \ac{DL} channel.
For reasons of comparability, we use either \ac{UL} or \ac{DL} channels as training data, i.e, either $\mathcal{H}_{4\times16}^{\text{UL,train}}$ or $\mathcal{H}_{4\times16}^{\text{DL,train}}$, to construct the codebooks with the procedure described in \Cref{sec:cb_design}.
Already with a codebook of $M=3$ bits, higher spectral efficiencies can be achieved as compared to the ``{\tt uni pow cov}'' transmit strategy. \\
\indent\emph{v}) With an increasing codebook size, the spectral efficiency increases and almost reaches the performance of the ``{\tt all train data \ac{UL}}'' or ``{\tt all train data \ac{DL}}'' transmit strategies, which constitute a natural upper bound on codebook based approaches with the respective \ac{UL} or \ac{DL} dataset, and which are computed as follows.
For each channel sample in $\mathcal{H}_{4\times16}^{\text{UL,train}}$ or $\mathcal{H}_{4\times16}^{\text{DL,train}}$, the corresponding capacity achieving transmit covariance matrix is calculated and is used as a codebook entry.
Thus, in this case, we have a huge codebook with \( \Ttr = 10^4 \) entries. \\
%\red{This approach provides an upper bound for transmit strategies, which use training data to construct a codebook.}
\indent Remarkably, the considered codebooks show comparable performance in the two cases where either \ac{UL} or \ac{DL} training data is used for the codebook construction.
In particular, the boxplots in~\Cref{fig:16x4_Mbit_se} show very similar distributions of the spectral efficiencies of the corresponding \ac{UL}/\ac{DL} pairs.
Whereas the instantaneous \ac{UL} and \ac{DL} channel realizations may differ,
one may now reasonably conjecture that the totality of the collected \ac{UL} and \ac{DL} channels share relevant statistical properties.
This conjecture was in particular investigated in~\cite{utschick2021,rizzello2021} for a large range of frequency gaps between the UL and DL domain and corroborated with the help of hypothesis tests.

\subsection{Codebook Performance with Estimated Channels}

In this section, we investigate how the distribution of the spectral efficiencies is affected when a codebook entry is selected with the help of estimated channels.

In the pilot transmission phase, the \ac{DL} received signal is:
\begin{equation} \label{eq:noisy_obs}
    \mbY = \mbH \mbP + \mbN \in \C^{\Nrx \times n_p} ,
\end{equation}
where $n_p$ is the number of transmitted pilots, $\mbN = [\mbn_1, \mbn_2, \dots, \mbn_{n_p}] \in \C^{\Nrx \times n_p}$ collects noise samples with 
\begin{equation}
\mbn_p \sim \mathcal{N}_\C(\mathbf{0}, \sigma_n^2 \mbI_{\Nrx}), 
\end{equation}
$p \in \{1,2, \dots, n_p\} $,
and where the pilot matrix $\mbP$ is given by
{\makeatletter
    \def\tagform@#1{\maketag@@@{\normalsize(#1)\@@italiccorr}}
\makeatother
\scriptsize
\begin{equation}
    \mbP = \sqrt{\dfrac{\rho}{N_t}} 
    \begin{bmatrix} 
    1 & 1 & \dots & 1 \\ 
    1 & W_{n_p} & \dots & W_{n_p}^{n_p-1} \\
    \vdots & \vdots & & \vdots \\
    1 & W_{n_p}^{\Ntx-1} & \dots & W_{n_p}^{(\Ntx-1)(n_p-1)} 
    \end{bmatrix} \in \C^{\Ntx \times n_p},
\end{equation}
}%
with $W_{n_p} = e^{\text{j}2\pi/n_p}$.
The pilot matrix $\mbP$ is a submatrix of the \ac{DFT} matrix. If $ n_p \geq N_\text{tx} $, the \ac{LS} channel estimate in the pilot transmission phase can be obtained as $\mbH_{\text{LS}} = \mbY \mbP^\dagger$ with the pseudoinverse $\mbP^\dagger = \mbP^\herm (\mbP \mbP^\herm)^{-1}$.%\footnote{Otherwise, the abbrevation LS---by abuse of notation---still denotes the estimation based on the pseudoinverse for underdetermined systems of equations.}

%As an alternative to least squares, it is common to assume that channels at the considered frequencies allow for a sparse representation.
As an alternative to least squares, it is common to assume that channels exhibits a certain structure.
Using the vectorized channel $\mbh = \mathrm{vec}(\mbH)$, we express this as
%Additionally, we assume to have structure in our channels, and therefore approximate the vectorized channel $\mbh = \text{vec}(\mbH)$ by
$\mbh \approx \mbD \mbt$, where $\mbD = \mbD_{\text{rx}} \otimes \mbD_{\text{tx}}$ is a \textit{dictionary} with oversampled \ac{DFT} matrices $\mbD_{\text{rx}}$ and $\mbD_{\text{tx}}$ (cf., e.g., \cite{AlLeHe15}), because we have \acp{ULA} at the transmitter and receiver side.
A compressive sensing algorithm like \ac{OMP}~\cite{Gharavi} can now be used to obtain a sparse vector \( \mbt \), and the channel estimate then computes to \( \mathrm{vec}(\mbH_{\text{OMP}}) = \mbD \mbt \).
Since the sparsity order is not known but the algorithm's performance crucially depends on it, we use a genie-aided approach to obtain an upper bound on the performance.
To this end, we use the true channel to choose the optimal sparsity for \ac{OMP}.

\begin{figure}[tb]
\centering
\begin{tikzpicture}
\begin{axis}[
  legend style={at={(0,1)},anchor=north west},
  boxplot/draw direction=y,
  ylabel={\scriptsize{spectral efficiency [bits/c.u.]}},
  height=\boxplotheight,
  width=\boxplotwidth,
  title={\small{(a) \(\quad \Ntx = 16, \Nrx = 4, n_p = 16\)}},
  xmajorticks=false,
  y tick label style={
      font=\scriptsize,
      align=center,
  },
  cycle list={},
  y label style={at={(-0.09,0.5)}},
]

% uniform power allocation
\addplot+[
    TUMBeamerRed,
    boxplot prepared={
      median=8.349689860148676,
      upper quartile=14.612202291480003,
      lower quartile=3.2180369215243636,
      upper whisker=26.584417675500962,
      lower whisker=0.3489717488209078,
      every whisker/.style={\whiskerstyle},
      every median/.style={solid},
      every box/.style={fill=TUMBeamerRed,opacity = 0.25}
    },
    ]
    coordinates {};
    
% ls cb DL
\addplot+[
    TUMBeamerLightGreen,
    boxplot prepared={
      lower whisker=0.012243590209272659,
      lower quartile=5.330334598935082,
      median=11.579671409490125,
      upper quartile=18.491508322451985,
      upper whisker=32.876587814584816,
      every whisker/.style={\whiskerstyle},
      every median/.style={solid},
      every box/.style={fill=TUMBeamerLightGreen,opacity = 0.25}
    },
    ]
    coordinates {};

% ls cb UL
\addplot+[
    TUMBeamerLightGreen,
    boxplot prepared={
      lower whisker=0.006426489434989632,
      lower quartile=5.313246043569647,
      median=11.54845985874751,
      upper quartile=18.48039282867595,
      upper whisker=32.64132252121264,
      every whisker/.style={\whiskerstyle},
      every median/.style={solid},
      every box/.style={fill=TUMBeamerLightGreen,opacity = 0.5}
    },
    ]
    coordinates {};

% omp cb DL
\addplot+[
    TUMBlue,
    boxplot prepared={
      median=11.5318887546978,
      upper quartile=18.41703509049528,
      lower quartile=5.424622137993345,
      upper whisker=32.876587814584816,
      lower whisker=0.009557894478250442,
      every whisker/.style={\whiskerstyle},
      every median/.style={solid},
      every box/.style={fill=TUMBlue,opacity = 0.25}
    },
    ]
    coordinates {};

% omp cb UL
\addplot+[
    TUMBlue,
    boxplot prepared={
      lower whisker=0.01381567261318455,
      lower quartile=5.417898339885033,
      median=11.497429153858903,
      upper quartile=18.386227129516577,
      upper whisker=32.64132252116661,
      every whisker/.style={\whiskerstyle},
      every median/.style={solid},
      every box/.style={fill=TUMBlue,opacity = 0.5}
    },
    ]
    coordinates {};

% cb DL
\addplot+[
    TUMBeamerOrange,
    boxplot prepared={
      lower whisker=0.6290332623754722,
      lower quartile=5.724525871877813,
      median=11.780757947178925,
      upper quartile=18.588332083837482,
      upper whisker=32.876587814584816,
      every whisker/.style={\whiskerstyle},
      every median/.style={solid},
      every box/.style={fill=TUMBeamerOrange,opacity = 0.25}
    },
    ]
    coordinates {};

% cb UL
\addplot+[
    TUMBeamerOrange,
    boxplot prepared={
      lower whisker=0.6617899174777325,
      lower quartile=5.703155972937565,
      median=11.779116671236295,
      upper quartile=18.60336720374872,
      upper whisker=32.64132252121264,
      every whisker/.style={\whiskerstyle},
      every median/.style={solid},
      every box/.style={fill=TUMBeamerOrange,opacity = 0.5}
    },
    ]
    coordinates {};

% LS wf
\addplot+[
    TUMBeamerGreen,
    boxplot prepared={
      lower whisker=0.18198950063034575,
      lower quartile=5.439832921619351,
      median=13.158183095328992,
      upper quartile=21.246599840722077,
      upper whisker=34.471609975271136,
      every whisker/.style={\whiskerstyle},
      every median/.style={solid},
      every box/.style={fill=TUMBeamerGreen,opacity = 0.7}
    },
    ]
    coordinates {};

% omp wf
\addplot+[
    TUMBlue,
    boxplot prepared={
      lower whisker=0.0023783718695258513,
      lower quartile=5.998260923255056,
      median=13.216152635033973,
      upper quartile=21.025415140345036,
      upper whisker=34.43772003696838,
      every whisker/.style={\whiskerstyle},
      every median/.style={solid},
      every box/.style={fill=TUMBlue,opacity = 0.7}
    },
    ]
    coordinates {};

% uni pow H
\addplot+[
    TUMDarkGray,
    boxplot prepared={
      lower whisker=1.14042986285615,
      lower quartile=7.147132574728896,
      median=14.523214477953129,
      upper quartile=21.892799160837,
      upper whisker=34.5327375451216,
      every whisker/.style={\whiskerstyle},
      every median/.style={solid},
      every box/.style={fill=TUMDarkGray,opacity = 0.25}
    },
    ]
    coordinates {};

% H wf
\addplot+[
    TUMDarkGray,
    boxplot prepared={
      lower whisker=1.758560081435993,
      lower quartile=7.746208293061338,
      median=14.662980867253147,
      upper quartile=21.90918050179577,
      upper whisker=34.532747040316906,
      every whisker/.style={\whiskerstyle},
      every median/.style={solid},
      every box/.style={fill=TUMDarkGray,opacity = 0.7}
    },
    ]
    coordinates {};

% means
\addplot[line width=1,solid, mark=*, opacity=0.5, mark size=0.5pt]%,mark=square]
        table[x=x,y=mean] {16x4_6bit_se_means.txt};
%        \addlegendentry{\footnotesize{mean}};

\legend{,,,,,,,,,,,\scriptsize{mean}};

\end{axis}

\begin{axis}[
  yshift=-5.4cm,
  legend style={at={(0,1)},anchor=north west},
  boxplot/draw direction=y,
  ylabel={\scriptsize{spectral efficiency [bits/c.u.]}},
  height=\boxplotheight,
  width=\boxplotwidth,
  title={\small{(b) \(\quad \Ntx = 32, \Nrx = 16, n_p = 32\)}},
  extra y ticks={25,75},
  xtick={1,2,...,50},
  xticklabels={%
    \scriptsize{\pltUniPowCov},
    \scriptsize{\pltHlsDLcb},
    \scriptsize{\pltHlsULcb},
    \scriptsize{\pltHompDLcb},
    \scriptsize{\pltHompULcb},
    \scriptsize{\pltHDLcb},
    \scriptsize{\pltHULcb},
    \scriptsize{\pltHlsWfCov},
    \scriptsize{\pltHompWfCov},
    \scriptsize{\pltHeigsp},
    \scriptsize{\pltWfCov},
  },
  x tick label style={
      text width=1.9cm,
      font=\scriptsize,
      align=right,
      rotate=80
  },
  xtick style={draw=none},
  y tick label style={
      font=\scriptsize,
      align=center,
  },
  cycle list={},
  y label style={at={(-0.09,0.5)}},
]

% uniform power allocation
\addplot+[
    TUMBeamerRed,
    boxplot prepared={
      median=25.387939374733108,
      upper quartile=53.32038060444651,
      lower quartile=8.58295448555031,
      upper whisker=101.49878217940558,
      lower whisker=1.437425658134748,
      every whisker/.style={\whiskerstyle},
      every median/.style={solid},
      every box/.style={fill=TUMBeamerRed,opacity = 0.25},
    },
    ]
    coordinates {};

% ls cb DL
\addplot+[
    TUMBeamerLightGreen,
    boxplot prepared={
      lower whisker=0.8253907599285593,
      lower quartile=12.598785332801352,
      median=32.330215399061515,
      upper quartile=60.364305839378886,
      upper whisker=105.53560187992534,
      every whisker/.style={\whiskerstyle},
      every median/.style={solid},
      every box/.style={fill=TUMBeamerLightGreen,opacity = 0.25}
    },
    ]
    coordinates {};

% ls cb UL
\addplot+[
    TUMBeamerLightGreen,
    boxplot prepared={
      lower whisker=0.8164460255662255,
      lower quartile=12.492208289300542,
      median=32.411447516072066,
      upper quartile=60.367677581114144,
      upper whisker=104.64524189024347,
      every whisker/.style={\whiskerstyle},
      every median/.style={solid},
      every box/.style={fill=TUMBeamerLightGreen,opacity = 0.5}
    },
    ]
    coordinates {};

% omp cb DL
\addplot+[
    TUMBlue,
    boxplot prepared={
      median=32.48359720001373,
      upper quartile=60.27559166037951,
      lower quartile=13.307120712342856,
      upper whisker=105.53560187992534,
      lower whisker=0.046640875365867834,
      every whisker/.style={\whiskerstyle},
      every median/.style={solid},
      every box/.style={fill=TUMBlue,opacity = 0.25}
    },
    ]
    coordinates {};

% omp cb UL
\addplot+[
    TUMBlue,
    boxplot prepared={
      lower whisker=0.06222548447869398,
      lower quartile=13.206786580957864,
      median=32.4191641054707,
      upper quartile=60.308569518968525,
      upper whisker=104.64524189024347,
      every whisker/.style={\whiskerstyle},
      every median/.style={solid},
      every box/.style={fill=TUMBlue,opacity = 0.5}
    },
    ]
    coordinates {};

% cb DL
\addplot+[
    TUMBeamerOrange,
    boxplot prepared={
      lower whisker=2.209257325758257,
      lower quartile=13.530671182541207,
      median=32.65451452155429,
      upper quartile=60.39989684726338,
      upper whisker=105.53560187992534,
      every whisker/.style={\whiskerstyle},
      every median/.style={solid},
      every box/.style={fill=TUMBeamerOrange,opacity = 0.25}
    },
    ]
    coordinates {};

% cb UL
\addplot+[
    TUMBeamerOrange,
    boxplot prepared={
      lower whisker=2.21960382249268,
      lower quartile=13.46197986728402,
      median=32.65923270534802,
      upper quartile=60.38374685082575,
      upper whisker=104.64524189024347,
      every whisker/.style={\whiskerstyle},
      every median/.style={solid},
      every box/.style={fill=TUMBeamerOrange,opacity = 0.5}
    },
    ]
    coordinates {};

% LS wf
\addplot+[
    TUMBeamerGreen,
    boxplot prepared={
      lower whisker=1.9151892405217865,
      lower quartile=11.999705364236561,
      median=33.29473648813462,
      upper quartile=65.04472974887213,
      upper whisker=116.9452751614595,
      every whisker/.style={\whiskerstyle},
      every median/.style={solid},
      every box/.style={fill=TUMBeamerGreen,opacity = 0.7}
    },
    ]
    coordinates {};

% omp wf
\addplot+[
    TUMBlue,
    boxplot prepared={
      lower whisker=0.04968033616403606,
      lower quartile=14.026516849463931,
      median=34.46663540705349,
      upper quartile=64.00193553891208,
      upper whisker=113.65343175966079,
      every whisker/.style={\whiskerstyle},
      every median/.style={solid},
      every box/.style={fill=TUMBlue,opacity = 0.7}
    },
    ]
    coordinates {};

% uni pow H
\addplot+[
    TUMDarkGray,
    boxplot prepared={
      lower whisker=2.4174299441524463,
      lower quartile=13.200566153465147,
      median=34.33401576811411,
      upper quartile=66.3655731885069,
      upper whisker=117.22652311102945,
      every whisker/.style={\whiskerstyle},
      every median/.style={solid},
      every box/.style={fill=TUMDarkGray,opacity = 0.25}
    },
    ]
    coordinates {};

% H wf
\addplot+[
    TUMDarkGray,
    boxplot prepared={
      lower whisker=5.681235011893994,
      lower quartile=17.897127901003678,
      median=37.85656705797946,
      upper quartile=67.43364185828658,
      upper whisker=117.22877740635688,
      every whisker/.style={\whiskerstyle},
      every median/.style={solid},
      every box/.style={fill=TUMDarkGray,opacity = 0.7}
    },
    ]
    coordinates {};

% means
\addplot[line width=1,solid, mark=*, opacity=0.5, mark size=0.5pt]%,mark=square]
        table[x=x,y=mean] {32x16_6bit_se_means.txt};
%        \addlegendentry{\footnotesize{mean}};

\legend{,,,,,,,,,,,\scriptsize{mean}};

\end{axis}
\end{tikzpicture}
\vspace{-0.2cm}
\caption{Spectral efficiencies of different transmit strategies in two different scenarios. Codebooks with \( M = 6 \) bits are either constructed using UL or DL training data: ``{\tt UL cb}'' or ``{\tt DL cb}''. The LS and OMP channel estimates are \( \mbH_{\text{LS}} \) and \( \mbH_{\text{OMP}} \), and \( \mbH \) refers to the true DL channel. The ``{\tt wf}'' labels correspond to transmission with water-filling covariance matrices where complete channel knowledge would be necessary.}
\label{fig:6bit_se}
\end{figure}

\Cref{fig:6bit_se}(a) shows the spectral efficiency distributions of the scenario with $\Ntx=16$ and $\Nrx=4$ again.
The codebooks are the same as in the previous section ($M=6$)---trained using either $\mathcal{H}_{4\times16}^{\text{UL,train}}$ or $\mathcal{H}_{4\times16}^{\text{DL,train}}$---, and the same evaluation set $\mathcal{H}_{4\times 16}^{\text{DL,eval}}$ is used. \\
\indent\emph{i}) The boxplots labeled ``\( \mbH \), {\tt DL cb}'' or ``\( \mbH \), {\tt UL cb}'' select the feedback index in~\eqref{eq:codebook_index_selection} using the true \ac{DL} channels \( \mbH \), i.e., assuming perfect \ac{CSI}.
Consequently, this is an upper bound for index selection based on channel estimates.\\
\indent\emph{ii}) After receiving $n_p = 16$ pilots, the \ac{MT} estimates the \ac{DL} channel and uses this estimated channel to select a codebook entry via~\eqref{eq:codebook_index_selection}.
The corresponding boxplots are denoted by ``$\mbH_{\text{LS}}$, {\tt DL cb}'' or ``$\mbH_{\text{LS}}$, {\tt UL cb}'' and ``{$\mbH_{\text{OMP}}$, {\tt DL cb}}'' or ``{$\mbH_{\text{OMP}}$, {\tt UL cb}}'', respectively, depending on whether \ac{LS} or \ac{OMP} channel estimates are used and whether the codebooks are constructed with \ac{UL} or \ac{DL} data.
It can be seen that irrespective of whether we have constructed the codebook with \ac{DL} or \ac{UL} training data, the distributions of the spectral efficiencies hardly differ at all.
\ac{OMP} seems to perform slightly better than \ac{LS}.\\
%We again have as a performance upper bound, the spectral efficiencies obtained using the optimal transmit strategy, denoted by ``optimal'' and the ``uni pow eigsp chan'' transmit strategy, where a transmit covariance matrix is calculated by allocating equal power on the eigenspace of the channel.
%As a lower bound we have the ``uni pow'' approach.
\indent\emph{iii}) Further, we depict approaches (``$\mbH_{\text{LS}}$, {\tt wf cov}'' and ``$\mbH_{\text{OMP}}$, {\tt wf cov}''), where we use the estimated channel to calculate a transmit covariance matrix with water-filling.
%by decomposing the channel estimate into parallel streams and employing waterfilling. 
These approaches are practically infeasible, since the \ac{BS} would require knowledge of the estimated channels.
\ac{OMP} is again slightly better than \ac{LS}.\\
\indent\emph{iv}) As a performance upper bound, we depict (``$\mbH$, {\tt wf cov}''), where the optimal transmit strategy is employed.
The approach ``$\mbH$, {\tt uni pow eigsp}'' is again the transmit strategy, where a transmit covariance matrix is calculated by allocating equal power on the eigenvectors of the channel. Recall, that these two approaches are also infeasible because the \ac{BS} would require full knowledge of the \ac{DL} channel.

Similar observations can be made in \Cref{fig:6bit_se}(b) where we have a different scenario: $\Ntx=32$, $\Nrx=16$, and $n_p=32$.
As before, we split the data into training sets $\mathcal{H}_{16\times32}^{\text{UL,train}}$ and $\mathcal{H}_{16\times32}^{\text{DL,train}}$ with $10^4$ samples each and an evaluation set $\mathcal{H}_{16\times32}^{\text{DL,eval}}$.

\section{Deep Neural Network for Feedback Encoding}

Instead of using the two stage process of first estimating a channel and then selecting a codebook entry via~\eqref{eq:codebook_index_selection}, we propose to find a function which directly maps from the noisy observations $\mbY$ to the feedback index $k^\star$:
\begin{equation}
    f: \mathbb{C}^{\Nrx \times n_p } \rightarrow \{1, \cdots, 2^M\}, \hspace{0.25cm} \mbY \mapsto f(\mbY) = k^\star.
\end{equation}
This is readily interpreted as a classification task.
In particular, we propose to use a \ac{DNN} \( f_{\text{DNN}} \) to approximate the function $f$ and, thus, to perform the classification task.
%\begin{equation}
%    f(\mbY) \approx f_{\text{DNN}}(\mbY) 
%\end{equation}

The codebooks ($M=6$) are the same as in the previous section---trained using either $\mathcal{H}_{\Nrx \times \Ntx}^{\text{UL,train}}$ or $\mathcal{H}_{\Nrx \times \Ntx}^{\text{DL,train}}$ for the two considered scenarios.
%with $\Ntx=16$ and $\Nrx=4$ or, $\Ntx=32$ and $\Nrx=16$.
We further use the four training data sets to generate labeled data for four different \ac{DNN} approaches: either \ac{UL} or \ac{DL} with either $(\Ntx, \Nrx)=(16, 4)$ or $(\Ntx, \Nrx)=(32, 16)$.
The labels are given by the optimal codebook indices determined via~\eqref{eq:codebook_index_selection}.
%To this end, we determine the codebook entry index, which maximizes the spectral efficiency, see \eqref{eq:codebook_index_selection}. 
%These indices are now our output labels.
The input/output pairs $\{(\mbY_{n},k^\star_n)\}_{n=1}^{\Ttr}$ then form the \ac{DNN} training data sets.
%The input labels are noisy observations obtained with \eqref{eq:noisy_obs} for a specific number of pilots $n_p$. 
%For each of the four configurations the labeled training set is then: $\{(\mbY_{i},k^*_i)\}_{n=1}^{\Ttr}$.

%The remaining data in $\mathcal{H}_{\Nrx \times \Ntx}^{\text{UL}}$ or $\mathcal{H}_{\Nrx \times \Ntx}^{\text{DL}}$ for the scenarios $(\Ntx, \Nrx)=(16, 4)$ or $(\Ntx, \Nrx)=(32, 16)$ are split into a validation part $\mathcal{H}_{\Nrx \times \Ntx}^{\text{UL,val}}$ or $\mathcal{H}_{\Nrx \times \Ntx}^{\text{DL,val}}$ with $\Tval = |\mathcal{H}_{\Nrx \times \Ntx}^{\text{UL,val}}| = |\mathcal{H}_{\Nrx \times \Ntx}^{\text{DL,val}}| = 2500$ and the remaining data constitute our test part $\mathcal{H}_{\Nrx \times \Ntx}^{\text{UL,test}}$ or $\mathcal{H}_{\Nrx \times \Ntx}^{\text{DL,test}}$.
Of the remaining data in $\mathcal{H}_{\Nrx \times \Ntx}^{\text{UL}}$ or $\mathcal{H}_{\Nrx \times \Ntx}^{\text{DL}}$ for the two scenarios, $2500$ samples are split off to form four validation data sets, and, finally, the rest is used for test sets.
Thus, the test sets consist of \(6273\) samples for \( (\Ntx, \Nrx) = (16, 4) \) and \(3648\) samples for \( (\Ntx, \Nrx) = (32, 16) \).

%Note that, $\Ttest = |\mathcal{H}_{\Nrx \times \Ntx}^{\text{UL,test}}| = |\mathcal{H}_{\Nrx \times \Ntx}^{\text{DL,test}}|$ is $6273$ for the scenario with $(\Ntx, \Nrx)=(16, 4)$ and is $3648$ for the scenario with $(\Ntx, \Nrx)=(32, 16)$.

\subsection{DNN Structure and Training Procedure}

The considered \ac{DNN} has two input ``channels'': one for the real part \( \Re(\mbY) \) of the observation \( \mbY \), and one for the imaginary part \( \Im(\mbY) \).
In a first step, these two parts are normalized w.r.t. the Frobenius norm.
We employ random search~\cite{BeBe12} to determine the hyperparameters (explained next) of the \ac{DNN}.

%The input to the \ac{DNN} is the noisy obseravtion $\mbY$ split into its real and imaginary part, and subsequently the real and imaginary parts are normalized by their respective Frobenius norms, i.e., our normalization is sample specific and does not depend on any precalculated normalization constants (such as means or variances of the respective data sets). Accordingly, we have two input ``channels'' into our \ac{DNN}.
%The network architecture and the training procedure is based on random search~\cite{BeBe12}.
The first modules of the \ac{DNN} are convolutional modules, which consist of a convolutional layer, a batch normalization layer, and an activation function.
We have $D_{\text{CM}}$ such modules, where $D_{\text{CM}}$ is randomly chosen from $[8, 20]$.
Each of the convolutional layers has $D_{\text{K}}$ kernels, where $D_{\text{K}}$ is randomly chosen from $[32, 64]$.
In the setting \((\Ntx, \Nrx) = (16, 4) \), we then flatten the features and obtain a vector of size $ N_f = D_{\text{K}} \Nrx n_p$.
In the setting \( (\Ntx, \Nrx) = (32, 16) \), two-dimensional max pooling by a factor of two is applied prior to flattening, yielding a \( N_f = D_{\text{K}} \frac{\Nrx}{2}  \frac{n_p}{2} \)-dimensional vector.
%For our second configuration, i.e., $(\Ntx, \Nrx)=(32, 16)$, prior to flattening we apply two dimensional max pooling by a factor of two and thus obtain a vector of size $D_{\text{K}} \cdot \Nrx/2 \cdot n_p/2$.
Subsequently, we have fully connected layers---each followed by batch normalization and activation function---, decreasing the dimension from \( N_f \) to $ 512 \rightarrow 256 \rightarrow 128 \rightarrow 64 \rightarrow 64 = 2^M$.

The loss function is cross-entropy.
We train for $300$ epochs with a $5$ epochs early stopping criterion.
%The number of epochs is set to $300$, with an early stopping criterion of $5$ epochs.
The activation function is the same in each layer, but randomly chosen from $ \{\text{ReLu, sigmoid, PReLU, Leaky ReLU, tanh, swish}\}$. Further random parameters are:
%the weight initialization \( \in \{ \text{xavier normal},\text{ xavier uniform}, \text{kaiming uniform}, \text{kaiming normal}, \text{normal}, \text{uniform} \} \) with standard parameters,
batch size $\in [20,1000]$, learning rate $ \in [10^{-6}, 10^{-3}]$, $L_1$ regularization $\in [10^{-6}, 10^{-3}]$, $L_2$ regularization $\in [10^{-6}, 10^{-3}]$, and exponential learning rate decay $\in [0.94, 1]$.
The optimizer is Adam \cite{Kingma}. 
We run $100$ random searches and pick out the best \ac{DNN}.
%out the \ac{DNN}, which achieves the best average spectral efficiency on the respective test set.

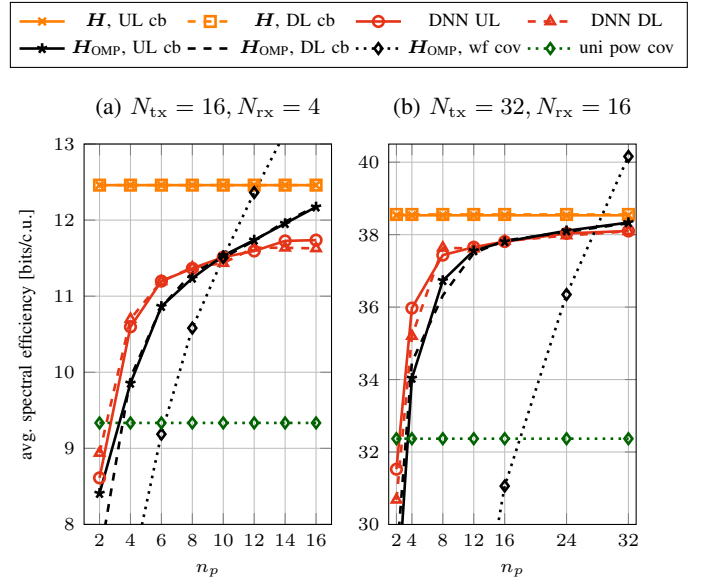
\begin{figure}[tb]
    \centering
    \begin{tikzpicture}
        \begin{axis}[
            height=\smallplotheight,
            width=\smallplotwidth,
            legend pos= north east,
            legend style={font=\scriptsize, at={(-0.3, 1.2)}, anchor=south west, legend columns=4},
            label style={font=\scriptsize},
            tick label style={font=\scriptsize},
            title={\small{(a) \(\Ntx = 16, \Nrx = 4\)}},
            xmin=1,
            xmax=17,
            ymin=8,
            ymax=13,
            xlabel={$n_p$},
            ylabel={avg. spectral efficiency [bits/c.u.]},
            xtick={2,4,6,8,10,12,14,16},
            grid=both,
        ]
            %\addplot[line width=1,TUMBlue,solid,mark=x,mark size=2pt]
            \addplot[cbUL]
                table[x=pilots,y=cb_UL] {plotoverpilots_16x4_omp.txt};
                \addlegendentry{$\mbH$, UL cb};
            \addplot[cbDL]
                table[x=pilots,y=cb_DL] {plotoverpilots_16x4_omp.txt};
                \addlegendentry{$\mbH$, DL cb};
            \addplot[cnnUL]
                table[x=pilots,y=CNN_UL] {plotoverpilots_16x4_omp.txt};
                \addlegendentry{DNN UL};
            \addplot[cnnDL]
                table[x=pilots,y=CNN_DL] {plotoverpilots_16x4_omp.txt};
                \addlegendentry{DNN DL};
            \addplot[hompUL]
                table[x=pilots,y=H_OMP_cb_UL] {plotoverpilots_16x4_omp.txt};
                \addlegendentry{$\mbH_{\text{OMP}}$, UL cb};
            \addplot[hompDL]
                table[x=pilots,y=H_OMP_cb_DL] {plotoverpilots_16x4_omp.txt};
                \addlegendentry{$\mbH_{\text{OMP}}$, DL cb};
            %\addplot[hlsUL]
            %    table[x=pilots,y=H_LS_cb_UL] {plotoverpilots_16x4_omp.txt};
            %    \addlegendentry{$\mbH_{\text{LS}}$, UL cb};
            %\addplot[hlsDL]
            %    table[x=pilots,y=H_LS_cb_DL] {plotoverpilots_16x4_omp.txt};
            %    \addlegendentry{$\mbH_{\text{LS}}$, DL cb};
            \addplot[hompWF]
                table[x=pilots,y=H_OMP_wf] {plotoverpilots_16x4_omp.txt};
                \addlegendentry{$\mbH_{\text{OMP}}$, wf cov};
            \addplot[unipow]
                table[x=pilots,y=uniform] {plotoverpilots_16x4_omp.txt};
                \addlegendentry{uni pow cov};
            %\addplot[hlsWF]
            %    table[x=pilots,y=H_LS_wf] {plotoverpilots_16x4_omp.txt};
            %    \addlegendentry{$\mbH_{\text{LS}}$, wf cov};
        \end{axis}    
    
        \begin{axis}[
            height=\smallplotheight,
            width=\smallplotwidth,
            legend pos= north east,
            legend style={font=\scriptsize, at={(0.0, 1.2)}, anchor=south west, legend columns=4},
            label style={font=\scriptsize},
            tick label style={font=\scriptsize},
            title={\small{(b) \(\Ntx = 32, \Nrx = 16\)}},
            xmin=1,
            xmax=33,
            ymin=30,
            ymax=40.5,
            xlabel={$n_p$},
            xtick={2,4,8,12,16,24,32},
            grid=both,
            xshift=4.05cm,
        ]
            \addplot[cbUL]
                table[x=pilots,y=cb_UL] {plotoverpilots_32x16_omp.txt};
                %\addlegendentry{cb UL};
            \addplot[cbDL]
                table[x=pilots,y=cb_DL] {plotoverpilots_32x16_omp.txt};
                %\addlegendentry{cb DL};
            \addplot[cnnUL]
                table[x=pilots,y=CNN_UL] {plotoverpilots_32x16_omp.txt};
                %\addlegendentry{CNN UL};
            \addplot[cnnDL]
                table[x=pilots,y=CNN_DL] {plotoverpilots_32x16_omp.txt};
                %\addlegendentry{CNN DL};
            \addplot[hompUL]
                table[x=pilots,y=H_OMP_cb_UL] {plotoverpilots_32x16_omp.txt};
                %\addlegendentry{$\mbH_{\text{OMP}}$ UL};
            \addplot[hompDL]
                table[x=pilots,y=H_OMP_cb_DL] {plotoverpilots_32x16_omp.txt};
                %\addlegendentry{$\mbH_{\text{OMP}}$ DL};
            %\addplot[hlsUL]
            %    table[x=pilots,y=H_LS_cb_UL] {plotoverpilots_32x16_omp.txt};
                %\addlegendentry{$\mbH_{\text{LS}}$ UL};
            %\addplot[hlsDL]
            %    table[x=pilots,y=H_LS_cb_DL] {plotoverpilots_32x16_omp.txt};
                %\addlegendentry{$\mbH_{\text{LS}}$ DL};
            \addplot[hompWF]
                table[x=pilots,y=H_OMP_wf] {plotoverpilots_32x16_omp.txt};
                %\addlegendentry{$\mbH_{\text{OMP}}$ wf};
            \addplot[unipow]
                table[x=pilots,y=uniform] {plotoverpilots_32x16_omp.txt};
                %\addlegendentry{uni pow};
            %\addplot[hlsWF]
            %    table[x=pilots,y=H_LS_wf] {plotoverpilots_32x16_omp.txt};
                %\addlegendentry{$\mbH_{\text{LS}}$ wf};
        \end{axis}
    \end{tikzpicture}
    \caption{Average spectral efficiencies in two different scenarios. Codebooks with $M=6$ bits are constructed with \ac{UL} or \ac{DL} data and a \ac{DNN} trained with \ac{UL} or \ac{DL} data is used for feedback encoding.}
    \label{fig:plotoverp_withomp}
\end{figure}

\Cref{fig:plotoverp_withomp} depicts spectral efficiencies over the number of pilots \( n_p \leq \Ntx \):\\
%In \Cref{fig:plotoverp_withomp}(a) and (b) we depict the spectral efficiencies of the two scenarios $(\Ntx, \Nrx)=(16, 4)$ and $(\Ntx, \Nrx)=(32, 16)$, respectively, over the number of transmitted pilots $n_p$. 
%Note, that we consider less pilots than we have transmit antennas, i.e., $n_p < \Ntx$. 
\indent\emph{i}) The curves labeled ``$\mbH$, {\tt DL cb}'' and ``$\mbH$, {\tt UL cb}'' select the feedback index with~\eqref{eq:codebook_index_selection} using the true \ac{DL} channels \( \mbH \) and thus represent an upper bound for the curves labeled ``$\mbH_{\text{OMP}}$, {\tt DL cb}'' and ``$\mbH_{\text{OMP}}$, {\tt UL cb}'', where a codebook entry---constructed with \ac{DL} or \ac{UL} training data---is selected via~\eqref{eq:codebook_index_selection} using \ac{OMP} channel estimates.
%One explanation may be that \ac{OMP} uses its dictionary to exploit structure in the channel.
%We can see, that the \ac{OMP} based approach performs better as compared to the \ac{LS} approach, especially when only a few pilots are transmitted, i.e., with the \ac{OMP} based approach the structure which is given in the channel can be exploited, especially in the scenario with $(\Ntx, \Nrx)=(32, 16)$ (\Cref{fig:plotoverp_withomp}(b)).
There is again not much difference between the codebooks constructed using \ac{UL} or \ac{DL} data.
Interestingly, the curve ``$\mbH_{\text{OMP}}$, {\tt wf cov}'', where we use an \ac{OMP} channel estimate to calculate a transmit covariance matrix with water-filling, is worse than the codebook based approaches ``$\mbH_{\text{OMP}}$, {\tt DL cb}'' and ``$\mbH_{\text{OMP}}$, {\tt UL cb}'', if only a few pilots are transmitted.
In contrast, the use of a codebook with pre-computed transmit covariance matrices proves advantageous when only a small number of pilots are transmitted. Recall, that the ``$\mbH_{\text{OMP}}$, {\tt wf cov}'' approach is practically infeasible, since the \ac{BS} would require knowledge of the estimated channel.\\
\indent\emph{ii}) The proposed \ac{DNN} approach is labeled ``{\tt DNN DL}'' and ``{\tt DNN UL}'', depending on whether \ac{DL} or \ac{UL} training data has been used.
%The proposed \ac{DNN} approach, where we train the neural network at the \ac{BS} and then offload the weights and biases to the \ac{MT} is labeled as ``DNN DL/UL'' depending on whether we have used \ac{UL} or \ac{DL} data during training.
Since the \ac{DNN} directly maps the observations to a feedback index $k^\star$, the \ac{MT} does not need to know the codebook, which is in contrast to the other approaches.
%Recall, that the \ac{DNN} directly maps the observations to a feedback index $k^\star$.
%Thus, the \ac{MT} does not need to know the codebook in order to select a feedback index $k^\star$, as compared to ``$\mbH_{\text{LS/OMP}}$, DL/UL cb'', where the \ac{MT} needs the codebook.
\Cref{fig:plotoverp_withomp} shows that for small and moderate numbers of pilots, the \ac{DNN} outperforms the approaches ``$\mbH_{\text{OMP}}$, {\tt DL cb}'' and ``$\mbH_{\text{OMP}}$, {\tt UL cb}'', where first a channel is estimated and then a feedback index is chosen from the codebook via~\eqref{eq:codebook_index_selection}.
Remarkably, the \acp{DNN} trained with \ac{UL} or \ac{DL} data perform equally well.

\section{Conclusion}

We showed that in \ac{FDD} systems, we can learn an adaptive codebook directly at the \ac{BS} by gathering \ac{UL} \ac{CSI}.
Further, a \ac{DNN} feedback encoder can also be trained at the \ac{BS} with the help of the codebook and \ac{UL} \ac{CSI}.
The feedback encoder's weights and biases can be offloaded to all \acp{MT} within the coverage area of the \ac{BS}. 
A \ac{MT} can then directly select a feedback index from noisy observations via this \ac{DNN} such that no channel estimation or knowledge of the codebook are necessary. The overall concept and the fact that both the codebooks and the feedback classifiers are learned exclusively at the BS makes the use of many different adaptive codebooks in the cell conceivable.

\bibliographystyle{IEEEtran}
\bibliography{IEEEabrv,biblio}

\end{document}